\documentclass[12pt]{article}

\usepackage{amssymb}
\usepackage{amsmath}
\usepackage{amscd}
\usepackage{latexsym}
\usepackage{graphicx}

\usepackage{caption}

\usepackage{cite}

\newcommand{\be}{\begin{equation}}
\newcommand{\ee}{\end{equation}}
\newcommand{\Dlt}{\Delta}
\newcommand{\dlt}{\delta}
\newcommand{\prt}{\partial}
\newcommand{\br}{{\bf r}}

\newcommand{\bj}{{\bf j}}

\newcommand{\bt}{\beta}
\newcommand{\vp}{\varphi}

\newcommand{\al}{\alpha}

\newcommand{\om}{\omega}
\newcommand{\Om}{\Omega}

\newcommand{\dgr}{\dagger}

\newcommand{\rgl}{\rangle}
\newcommand{\lgl}{\langle}

\begin{document}

\begin{center}
 
{\Large{\bf Atom Optics with Cold Bosons} \\ [5mm]

V.I. Yukalov$^{1}$ and E.P. Yukalova$^{2}$}  \\ [3mm]

{\it
$^1$Bogolubov Laboratory of Theoretical Physics, \\
Joint Institute for Nuclear Research, Dubna 141980, Russia \\ [2mm]

$^2$Laboratory of Information Technologies, \\
Joint Institute for Nuclear Research, Dubna 141980, Russia } \\ [3mm]

{\bf E-mails}: {\it yukalov@theor.jinr.ru}, ~~ {\it yukalova@theor.jinr.ru} \\ [2mm]

{\it corresponding author}: V.I. Yukalov

\end{center}

\vskip 1cm

\begin{abstract}

Trapped bosonic atoms can be cooled down to temperatures where the atomic cloud
experiences Bose-Einstein condensation. Almost all atoms in a dilute gaseous 
system can be Bose-condensed, which implies that this system is in a coherent 
state. The coherent atomic system enjoys many properties typical of coherent 
optical systems. It is possible to generate different condensate coherent modes 
similarly to the generation of optical modes. Several effects can be observed,
such as interference patterns, interference current, Rabi oscillations, harmonic 
generation, parametric conversion, Ramsey fringes, mode locking, dynamic transition
between Rabi and Josephson regimes, and atomic squeezing. 

\end{abstract}

\vskip 2mm


\newpage

\section{Introduction}

The term {\it Atom Optics} refers to phenomena and techniques exploiting wave properties 
of neutral atoms \cite{Adams_1}. Typical experiments employ cold, slowly moving neutral 
atoms, experiencing effects similar to photon beams. For instance, like optical beams, 
the atomic beams may exhibit diffraction and interference, and can be focused with a 
Fresnel zone plate \cite{Doak_2}, or a concave atomic mirror \cite{Berkhout_3}. Cold 
atoms can be used in atom interferometers \cite{Cronin_4} and, generally, in atomtronics
\cite{Amico}. 

After the realization of the Bose-Einstein condensation in traps, there has happened 
a boost of interest to coherent effects in atom optics, since a Bose-condensed system 
is a coherent system, similar to coherent light \cite{Yukalov_5}. Coherent atomic states 
can be created in traps \cite{Yukalov_5,Parkins_6,Dalfovo_7,Courteille_8,Andersen_9,Yukalov_10,
Bongs_11,Yukalov_12,Posazhennikova_13,Yukalov_14,Proukakis_15,Yurovsky_16,Yukalov_17}, 
as well as in optical lattices \cite{Morsh_19,Moseley_20,Yukalov_21,Krutitsky_22,Yukalov_23}. 

Even more common properties between atomic and light optics have been discovered after 
it has been suggested \cite{Yukalov_24} that in traps it is possible to generate 
non-ground-state Bose-Einstein condensates. Since the Bose-Einstein condensed state is a 
coherent state \cite{Yukalov_25} that, because of atomic interactions, is described by 
a nonlinear equation, the non-ground state condensates have been called {\it nonlinear 
coherent modes} \cite{Yukalov_24}. The generation of these atomic modes is similar to 
the excitation of optical modes of an optical resonator \cite{Vokhnik}. The principal 
difference is the nonlinearity due to atomic interactions because of which the modes of
Bose condensate are termed {\it nonlinear}.  

In the present communication, we give an account of the properties of Bose condensates 
with nonlinear coherent modes, emphasizing those that are analogous to the properties
of optical systems. We show that Bose-condensed systems with nonlinear coherent modes
allow for the realization of many features that are so important in quantum information 
processing. Although the list of analogies between atom and photon optics is rather 
long, here we concentrate on those effects that are connected with the nonlinear coherent 
modes.

\section{Non-ground-state Bose-Einstein condensate}

First of all, let us explain what is a non-ground-state condensate and why it has become 
available for realization only after the observation of Bose condensation in traps.
In the standard definition, a Bose-Einstein condensation is the effect of a macroscopic
number of atoms piling down to the ground state level of a statistical system. In a uniform 
system, the ground-state energy level is not separated by a gap from excited states forming 
a continuum. Therefore there are no other separate levels except the ground-state one. 
There exist nonequilibrium condensates, but there cannot exist non-ground-state condensates 
on some other levels, since there are no such separated excited levels. 

The situation is different for atoms in a trap \cite{Yukalov_5}, where there exists a whole 
spectrum of discrete energy levels. Then, in addition to the lowest ground-state level 
corresponding to the usual Bose condensate, as in Fig. 1, there are many other discrete 
levels. Hence, if atoms could assemble on some level above the lowest one, as in Fig. 2, 
this would exactly be a non-ground-state condensate.

It is clear that such a non-ground-state condensate cannot occur in an equilibrium system.
An additional energy needs to be applied in order to accumulate atoms on a level whose
energy would be higher than that of the ground state. For instance, if one wishes to 
transfer atoms from a level with energy  $E_1$ to a level with a higher energy $E_2$, one
should apply, e.g., an external field 
\be
\label{1}
V(\br,t) = V_1(\br) \cos(\om t) + V_2(\br) \sin(\om t) \; ,
\ee
alternating with the frequency being in resonance with the transition frequency
\be
\label{2}
\om_{21} \equiv E_2 - E_1 \;   ,
\ee
the detuning from the resonance being small,
\be
\label{3}
\left| \; \frac{\Dlt}{\om} \; \right| \ll 1 \;, \qquad 
\Dlt \equiv \om - \om_{21} \;    .
\ee
Here and in what follows, we set the Planck constant $\hbar$ to one.          

Already at this initial stage, we can notice the similarity between the process of 
transferring from one energy level to another an electron in an atom and transferring
from one energy level to another a group of atoms in a trap. Both these cases deal 
with a kind of a two-level system. The difference is that a transferred atomic electron 
is alone, while the condensate consists of a large group of interacting atoms, which 
make the modes nonlinear.  

The nonlinear coherent modes are also termed topological, since the related spatial 
densities have different number of zeroes, as in Fig. 3.

\section{Condensate wave function}

To obtain an equation for the condensate wave function, let us start with the 
Heisenberg equation for the field operator
\be
\label{4}
i\; \frac{\prt}{\prt t} \; \psi(\br,t) = H[\;\psi\;] \psi(\br,t) \;   ,
\ee
with the operator Hamiltonian
\be
\label{5}
 H[\;\psi\;] = - \; \frac{\nabla^2}{2m} + U(\br,t) +
\int \psi^\dgr(\br') \; \Phi(\br - \br') \; \psi(\br') \; d\br' \; .
\ee
Here $U({\bf r},t)$ is an external potential and $\Phi({\bf r}- {\bf r'})$ 
is a particle interaction potential. 

At very low temperature and asymptotically weak interactions all atoms are assumed 
to be Bose condensed. Since a Bose-condensed system is coherent, the condensate wave 
function is a coherent state \cite{Yukalov_5,Yukalov_17,Yukalov_24}, defined as an 
eigenstate of the field operator,
\be
\label{6}
 \psi(\br,t) \; | \; \eta \; \rgl = \eta(\br,t) \; | \; \eta \; \rgl \;  .
\ee
Averaging the Heisenberg equation (\ref{4}) over the coherent state yields the equation 
for the condensate wave function
\be
\label{7}
 i\; \frac{\prt}{\prt t} \; \eta(\br,t) = H[\;\eta\;] \eta(\br,t) \;   ,
\ee
with the Hamiltonian
\be
\label{8}
H[\;\eta\;] = - \; \frac{\nabla^2}{2m} + U(\br,t) +
\int  \Phi(\br - \br') \; |\; \eta(\br',t) \; |^2\; d\br' \;  .
\ee

Describing the Bose-condensed system by equation (\ref{7}) corresponds to the 
coherent approximation, as far as all atoms are assumed to be condensed. Equation 
(\ref{7}) was advanced by Bogolubov \cite{Bogolubov_26} in 1949 in his well known 
book "Lectures on Quantum Statistics" that has been republished numerous times 
(see, e.g., \cite{Bogolubov_27,Bogolubov_28,Bogolubov_29}). A detailed analysis 
of this equation, with finding periodic and vortex solutions, was given in a 
series of papers by Gross \cite{Gross_30,Gross_31,Gross_32,Gross_33,Gross_34} 
(see also \cite{Wu_35,Pitaevskii_36}). By its mathematical structure, (\ref{7}) 
is a nonlinear Schr\"{o}dinger equation \cite{Malomed_37}. 

Since all atoms are assumed to be Bose condensed, it is convenient to pass to the 
condensate wave function
\be
\label{9}
\eta(\br,t) = \sqrt{N} \; \vp(\br,t)  
\ee
normalized to one,
$$
\int |\; \vp(\br,t) \; |^2\; d\br = 1 \;   .
$$

The total external field consists of two parts, a stationary trapping potential 
$U({\bf r})$ and an alternating potential $V({\bf r},t)$,
\be
\label{10}
U(\br,t) = U(\br) + V(\br,t) \;   .
\ee
Respectively, the Hamiltonian (\ref{8}) can be split into two terms
\be
\label{11}
 H[\; \eta\; ] = \hat H[\; \vp\; ]  + V(\br,t) \;  ,
\ee
with the first term being
\be
\label{12}
\hat H[\;\vp \;] = - \; \frac{\nabla^2}{2m} + U(\br) + N
\int  \Phi(\br - \br') \; |\; \vp(\br',t) \; |^2\; d\br' \;  .
\ee
Then Eq. (\ref{7}) takes the form
\be
\label{13}
 i\; \frac{\prt}{\prt t}\; \vp(\br,t)  = \left\{ 
\hat H [\; \vp \; ] + V(\br,t) \right\} \; \vp(\br,t) \; .
\ee

Coherent modes are the solutions to the eigenproblem
\be
\label{14}
\hat H[\; \vp_n\; ] \; \vp_n(\br) = E_n \; \vp_n(\br) 
\ee
characterized by the stationary energy levels $E_n$. The general solution to 
Eq. (\ref{13}) can be represented as an expansion over the coherent modes,
\be
\label{15}
\vp(\br,t) = \sum_n c_n(t) \; \vp_n(\br) \; e^{-iE_n t} \;   ,
\ee
where the coefficient functions $c_n(t)$ are slowly varying in time, as 
compared with the exponential function,
\be
\label{16}
 \frac{1}{E_n} \; \left| \; \frac{d c_n(t)}{d t} \; 
\right| \ll 1 \;  .
\ee 
Similar representations in optics are called slowly-varying amplitude approximation 
\cite{Allen_38,Mandel_39}. 

The quantity
\be
\label{17}
n_n(t) \equiv |\; c_n(t) \; |^2
\ee
defines the fractional mode population that satisfies the normalization condition
\be
\label{18}
 \sum_n n_n(t) = 1 \;  .
\ee

\section{Resonant mode generation}

Separate coherent modes can be excited from the ground state by means of resonant 
generation. The resonance implies that the frequency of the modulating field $\om$ 
is close to one of the transition frequencies
\be
\label{19}
\om_{mn} \equiv E_m - E_n  \;  ,
\ee
so that the detuning from the resonance be small,
\be
\label{20}
\left| \; \frac{\Dlt_{mn}}{\om} \; 
\right| \ll 1 \; , \qquad \Dlt_{mn} \equiv \om - \om_{mn} \;  .
\ee

Atoms in a trap are rather rarified, and their interactions can be represented by the 
local interaction potential
\be
\label{21}
\Phi(\br) = \Phi_0 \dlt(\br) \; , \qquad \Phi_0 \equiv 4\pi \; \frac{a_s}{m} \; ,
\ee
where $a_s$ is a scattering length. The process of the mode generation is characterized 
by two amplitudes, the interaction amplitude
\be
\label{22}
\al_{mn} \equiv N \Phi_0 \int |\; \vp_m(\br) \; |^2 \; 
\left\{ 2\;  |\; \vp_n(\br) \; |^2 - |\; \vp_m(\br) \; |^2 \right\} \; d\br
\ee
and the modulation-field amplitude
\be
\label{23}
 \bt_{mn} \equiv \int \vp_m^*(\br) \; [ \; V_1(\br) - i V_2(\br) \; ] \;
\vp_n(\br) \; d\br \;  .
\ee

When a higher mode, that can be numbered by $2$, is generated from the ground state, 
numbered by $1$, then substituting expansion (\ref{15}) into equation (\ref{13}) yields 
the equations for the coefficient functions
$$
i\; \frac{dc_1}{dt} = \al_{12} |\; c_2\; |^2 c_1 + 
\frac{1}{2} \; \bt_{12} c_2 e^{i\Dlt_{12} t} \; ,
$$
\be
\label{24}
i\; \frac{dc_2}{dt} = \al_{21} |\; c_1\; |^2 c_2 + 
\frac{1}{2} \; \bt_{12}^* c_1 e^{-i\Dlt_{12} t} \;    .
\ee

Similarly, it is possible to modulate the trap by two fields with the frequencies, 
say $\omega_A$ and $\omega_B$, that are close to two of the transition frequencies
\be
\label{25}
\om_{21} \equiv E_2 - E_1 \; , \qquad \om_{32} \equiv E_3 - E_2 \; , \qquad 
\om_{31} \equiv E_3 - E_1 \; .
\ee
Then there can exist three types of the mode generation:
$$
\om_A = \om_{21} \; , \qquad \om_B = \om_{32} \qquad ( cascade) \; ,
$$
$$
\om_A = \om_{31} \; , \qquad \om_B = \om_{32} \qquad ( \Lambda \; type) \; ,
$$
\be
\label{26}
\om_A = \om_{31} \; , \qquad \om_B = \om_{21} \qquad ( V \; type) \;   ,
\ee
depending on which levels are connected by resonance. 

In the case of two applied alternating fields, with two frequencies, the amplitude
functions are described by the equations
$$
i\; \frac{dc_1}{dt} = \left( \al_{12} |\; c_2\; |^2  + 
 \al_{13} |\; c_3\; |^2 \right) c_1 + f_1   \; ,
$$
$$
i\; \frac{dc_2}{dt} = \left( \al_{21} |\; c_1\; |^2  + 
\al_{23} |\; c_3\; |^2 \right) c_2 + f_2   \; ,
$$
\be
\label{27}
i\; \frac{dc_3}{dt} = \left( \al_{31} |\; c_1\; |^2  + 
\al_{32} |\; c_2\; |^2 \right) c_3 + f_3 \;  ,
\ee
in which the functions $f_j$ depend of the type of the generation scheme.  

Note that instead of modulating the trap, it is possible to modulate the scattering 
length by means of Feshbach resonance \cite{Timmermans_40},
\be
\label{28}
 a_s(B) = a_s \left( 1 - \; \frac{\Dlt B}{ B - B_{res}}  \right) \; ,
\ee
where $\Delta B$ is the resonance width and 
\be
\label{29}
B(t) = B_0 + b_1 \cos(\om t) + b_2 \sin(\om t )
\ee
is the alternating magnetic field. Then the interaction amplitude also becomes modulated,
\be
\label{30}
 \Phi_0(t) = 4\pi \; \frac{a_s(B)}{m} \;  .
\ee

In addition to resonance conditions (\ref{20}) and (\ref{26}), there can exist higher-order
resonances \cite{Yukalov_41,Yukalov_42}. For example, in the case of two modes, there can
occur harmonic generation, when a single modulating field is used, as is shown in Fig. 4,
and parametric conversion, when two modulated fields are employed, as is shown in Fig. 5.
More generally, in the two-mode case, there can exist multiple harmonic generation, under
the resonance condition
\be
\label{31}
 n\om = \om_{21} \qquad ( n = 1,2,3,\ldots ) \;  ,
\ee
and multiple parametric conversion, under the resonance condition
\be
\label{32}
 \sum_j ( \pm \om_j) = \om_{21} \;  .
\ee

\section{Matter-wave interferometry}

Bose condensed atoms correspond to coherent mater waves, because of which different effects,
typical of coherent beams, exist \cite{Yukalov_43}. 

\subsection{Interference patterns}

The density of atoms inside a trap,
\be
\label{33}
 \rho(\br,t) = \sum_n \rho_n(\br,t) + \rho_{int}(\br,t) \;  ,
\ee
is given by the sum of the mode densities
\be
\label{34}
\rho_n(\br,t) = N |\; c_n(t) \vp_n(\br) \; |^2
\ee
and the interference pattern
\be
\label{35}
\rho_{int}(\br,t) = N \sum_{m\neq n} c_m^*(t)\; c_n(t) \; \vp_m^*(\br)\;
 \vp_n(\br)\;  e^{i\om_{mn} t} \;  .
\ee

\subsection{Interference current}

Since the system with coherent modes is not equilibrium, there exists atomic current inside 
the trap,
\be
\label{36}
 \bj(\br,t) = \sum_n \bj_n(\br,t) + \bj_{int}(\br,t) \;  ,
\ee
consisting of the sum of the mode currents
\be
\label{37}
\bj_n(\br,t) = \frac{N}{m} \;
{\rm Im}\; |\; c_n(t) \; |^2 \; \vp_n^*(\br) \nabla\vp_n(\br) 
\ee
and the interference current, also called internal Josephson current,
\be
\label{38}
 \bj_{int}(\br,t) = \frac{N}{m} \; 
{\rm Im}\; \sum_{m\neq n} c_m^*(t) \; c_n(t) \; 
[\; \vp_m^*(\br) \nabla\vp_n(\br) \; ] e^{i\om_{mn} t} \;  .
\ee

\subsection{Rabi oscillations}

Similarly to the Rabi oscillations of two-level systems in optics \cite{Rabi_44}, the 
coherent two-mode populations oscillate according to the law \cite{Yukalov_25}
$$
n_1 = 1 - \; \frac{|\bt_{12}|^2}{\Om^2} \; \sin^2\left( \frac{\Om t}{2} \right) \; ,
$$
\be
\label{39}
 n_2 =  \frac{|\bt_{12}|^2}{\Om^2} \; \sin^2\left( \frac{\Om t}{2} \right) \;  ,
\ee
where the initial conditions
\be
\label{40}
c_1(0) = 1 \; , \qquad c_2(0) = 0 \; ,
\ee
are assumed, and where the effective Rabi frequency is given by the expression
\be
\label{41}
\Om^2 = [\; \Dlt\om + \al_{12} n_2 - 
\al_{21} n_1 \; ]^2 + |\;\bt_{12}\;|^2 \;   ,
\ee
with $\Delta \omega \equiv \omega - \omega_{21}$.

\subsection{Ramsey fringes}

Ramsey fringes \cite{Ramsey_45} describe the excited mode population $|c_2|^2$, after 
the action of two consecutive $\pi/2$ pulses of temporal length $\tau$, separated by 
a long time interval $T \gg \tau$, as a function of $T$. In our case, under the initial 
condition (\ref{40}), the excited mode population reads as
\be
\label{42}
 n_2(2\tau +T) = \frac{|\bt_{12}|^2}{\Om^2} \;\left[ \; 
\cos\left( \frac{\Om \tau}{2} \right) + 
\frac{\Dlt}{\Om} \; \sin\left( \frac{\Dlt T}{2} \right) \; \right] \; ,
\ee
with the effective detuning 
\be
\label{43}
\Dlt \equiv \Dlt \om + \al_{12} n_2 -  \al_{21} n_1  \; ,
\ee
where $\cos(\Omega \tau/2) =\sqrt{2}/2$. Despite that the effective detuning is 
a function of atomic population fractions $n_j$, the population fraction (\ref{42}) 
has the form typical of the Ramsey fringes in optics \cite{Ramsey_45}.

\subsection{Mode locking}

There is a range of parameters, when the mode populations are locked in the limited 
regions
\be
\label{44}
\frac{1}{2} < n_1 \leq 1 \; , \qquad 0 \leq n_2 < \frac{1}{2} \qquad
\left( |\bt_{12}| + \Dlt_{21} < \frac{1}{2} \; \al_{12} \right) \;   ,
\ee
which is called the Rabi regime. Here, for simplicity, we set $\alpha_{12} = \alpha_{21}$.
Outside the Rabi regime, these modes become unlocked and oscillate in the whole available 
region between $0$ and $1$,
\be
\label{45}
0 \leq n_j \leq 1 \qquad
\left( \bt_{12} + \Dlt_{21} > \frac{1}{2} \; \al_{12} \right) \;  ,
\ee
which is termed the Josephson regime. On the critical line
\be
\label{46}
  \bt_{12} + \Dlt_{21} = \frac{1}{2} \; \al_{12} \;  ,
\ee
where the regime changes, there occur critical fluctuations and critical phenomena, similar 
to those happening at phase transitions \cite{Yukalov_46}.

\section{Atomic squeezing}

By analogy with the squeezed light, there can exist atomic squeezing \cite{Yukalov_43}. 
To describe this effect, it is necessary to keep in mind trapped atoms, when, because 
of the finiteness of the system, the gauge symmetry is not yet broken and atoms are to 
be characterized by field operators. For the coherent modes, we have the field operators 
$a_n$ and $a_n^\dagger$ satisfying the Bose commutation relations and the averaging 
conditions
\be
\label{47}
\lgl \; a_m^\dgr a_n \; \rgl = N c_m^* c_n \;   .
\ee
In the case of two modes, one can introduce the pseudo-spin operators
\be
\label{48}
S^+ = a_2^\dgr a_1 \; , \qquad  S^- = a_1^\dgr a_2 \; , \qquad 
S^z = \frac{1}{2} \left( a_2^\dgr a_2 - a_1^\dgr a_1\right) \; , 
\ee
satisfying the standard spin algebra
$$
 [\; S^+ , \; S^- \; ] = 2S^z \; , \qquad  [\; S^z , \; S^\pm \; ] = \pm S^\pm \; .
$$
To obey the averaging conditions (\ref{47}), the pseudo-spin operators can be represented
as the sums
\be
\label{49}
S^\al = \sum_{i=1}^N S_i^\al \;  .
\ee

Generally, for two operators $\hat{A}$ and $\hat{B}$, there is the Heisenberg uncertainty 
relation
\be
\label{50}
{\rm var}(\hat A) \; {\rm var}(\hat B) \geq \frac{1}{4} \; | \; \lgl \;
[\; \hat A , \; \hat B \; ] \; \rgl \; |^2 \;  ,
\ee
in which the operator variance is
$$
{\rm var}(\hat A)  \equiv \lgl \; \hat A^+ \hat A \; \rgl - 
|\;  \lgl \;  \hat A \; \rgl \; |^2 .
$$

The squeezing factor of an operator $\hat{A}$ with respect to $\hat{B}$ is defined as
\be
\label{51}
 Q(\hat A, \hat B) \equiv \frac{2 {\rm var}(\hat A)}{| \; \lgl \;
[\; \hat A , \; \hat B \; ] \; \rgl \; | } \;  .
\ee
Respectively, the squeezing factor of an operator $\hat{B}$ with respect to $\hat{A}$ is
\be
\label{52}
Q(\hat B, \hat A) \equiv \frac{2 {\rm var}(\hat B)}{| \; \lgl \;
[\; \hat A , \; \hat B \; ] \; \rgl \; | } \;   .
\ee
Then the Heisenberg uncertainty relation can be written as
\be
\label{53}
  Q(\hat A, \hat B) \; Q(\hat B, \hat A) \geq 1 \;  .
\ee
One says that $\hat{A}$ is squeezed with respect to $\hat{B}$, if $Q(\hat{A},\hat{B})< 1$. 
The squeezing of one operator with respect to another means that the physical observable
corresponding to the first operator can be measured more precisely than the observable
corresponding to the second operator. The uncertainty relation in the form (\ref{53}) tells
us that if $\hat{A}$ is squeezed with respect to $\hat{B}$, then $\hat{B}$ is not squeezed 
with respect to $\hat{A}$,

In our case, we consider the operators $S^z$ and $S^{\pm}$, with the related squeezing 
factor  
\be
\label{54}
 Q(S^z, S^\pm) = \frac{2{\rm var}(S^z)}{| \; \lgl \; S^\pm \; \rgl \; | } \; .
\ee
It follows that 
\be
\label{55}
Q(S^z, S^\pm) = \sqrt{1 - s^2} \;   ,
\ee
where $s$ is the atomic population difference
\be
\label{56}
 s = \frac{2}{N} \; \lgl \; S^z \; \rgl = |\; c_2\; |^2 - |\; c_1 \; |^2 \;  .
\ee
Since $s \leq 1$, the squeezing factor is almost always less than one, hence the operator 
$S^z$ is almost always squeezed with respect to $S^{\pm}$. In physical parlance, this 
implies that atomic population difference practically always can be measured more precisely
than the atomic current proportional to $\langle S^{\pm} \rangle$.

\section{Conclusion}

Atom optics is a branch of physics studying matter-wave properties of atoms. A system of 
Bose condensed atoms in a trap allows for the creation of non-ground-state condensates, 
when the energy levels above the ground state can become macroscopically occupied. These 
excited energy levels are described by the eigenfunctions of the stationary nonlinear 
Schr\"{o}dinger operator. The corresponding atomic states are called nonlinear coherent 
modes. The Bose condensate with nonlinear coherent modes has many properties analogous 
to those of the finite-level atoms in optics, because of which there appears the whole 
new branch of atom optics dealing with the effects typical of light optics with finite-level
atoms. Here the description of such characteristic effects is given for resonant generation 
of two or several modes. Matter-wave interferometry based on Bose condensates with nonlinear 
coherent modes is described, including such effects as interference patterns, interference 
current, Rabi oscillations, Ramsey fringes, harmonic generation, parametric conversion, 
mode locking, dynamic transition between Rabi and Josephson regimes, and atomic 
squeezing. More details can be found in the review article \cite{Yukalov_47}.

\vskip 1cm
{\bf Acknowledgments}

\vskip 2mm
We are grateful to V.S. Bagnato for many discussions.

\vskip 5mm 

{\bf Funding}

\vskip 2mm
No funds, grants, or other support was received.

\vskip 5mm
{\bf Author contributions}

\vskip 2mm
All authors equally contributed to the paper.

\vskip 5mm

{\bf Conflict of interests}

\vskip 2mm
The authors have no conflicts of interests. 

\vskip 5mm

{\bf Financial interests}

\vskip 2mm
The authors declare they have no financial interests.

\newpage

\begin{center}
{\Large{\bf Figure Captions }}
\end{center}

\vskip 1cm
{\bf Figure 1}. Bose-Einstein condensation in a trap. Macroscopic occupation of
a ground-state energy level.

\vskip 1cm
{\bf Figure 2}. Non-ground-state condensate in a trap. Macroscopic occupation of
a non-ground-state energy level.

\vskip 1cm
{\bf Figure 3}. Spatial dependence for the density of a ground-state condensate and
of a non-ground-state mode.

\vskip 1cm
{\bf Figure 4}. Harmonic generation for a Bose condensate with nonlinear coherent modes.

\vskip 1cm
{\bf Figure 5}. Parametric conversion for a Bose condensate with nonlinear coherent modes.

\newpage

\begin{figure}[ht]
\begin{center}
\includegraphics[width=6cm]{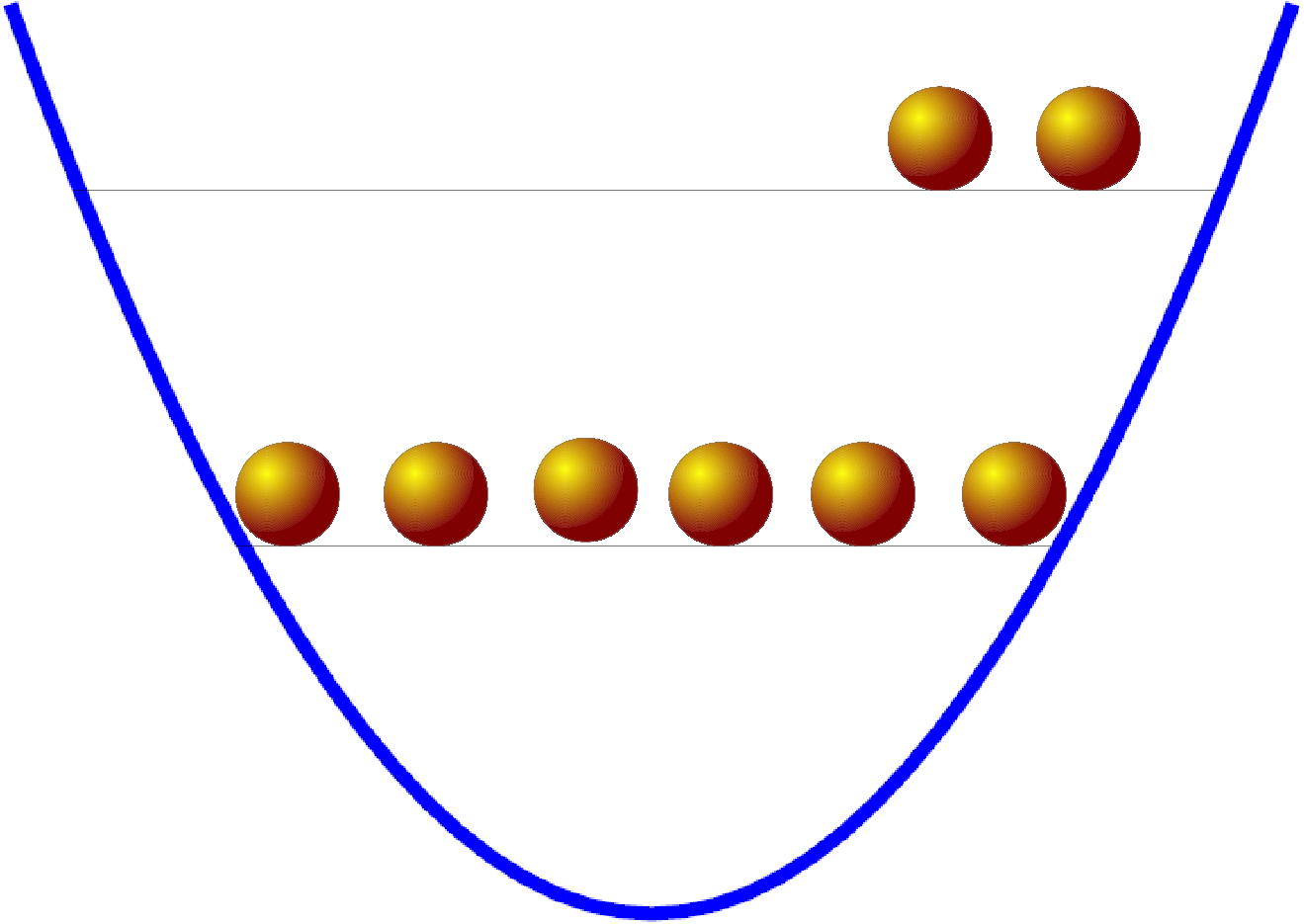}  
\end{center}
\label{Fig.1}
\end{figure}
{\bf Fig. 1}

\vskip 1cm
\begin{figure}[ht]
\begin{center}
\includegraphics[width=6cm]{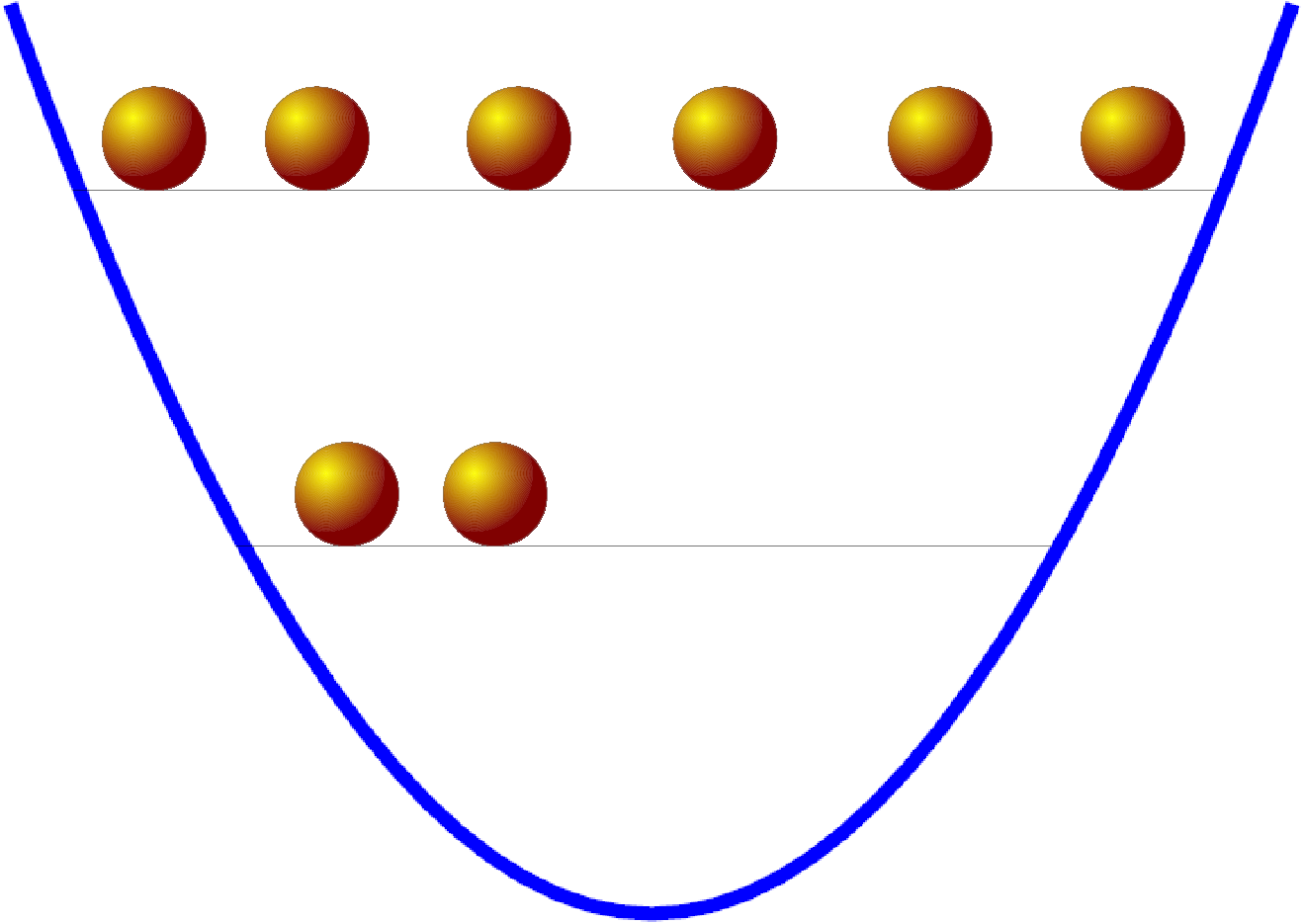}  
\end{center}
\label{Fig.2}
\end{figure}
{\bf Fig. 2}

\newpage
\begin{figure}[ht]
\begin{center}
\includegraphics[width=10cm]{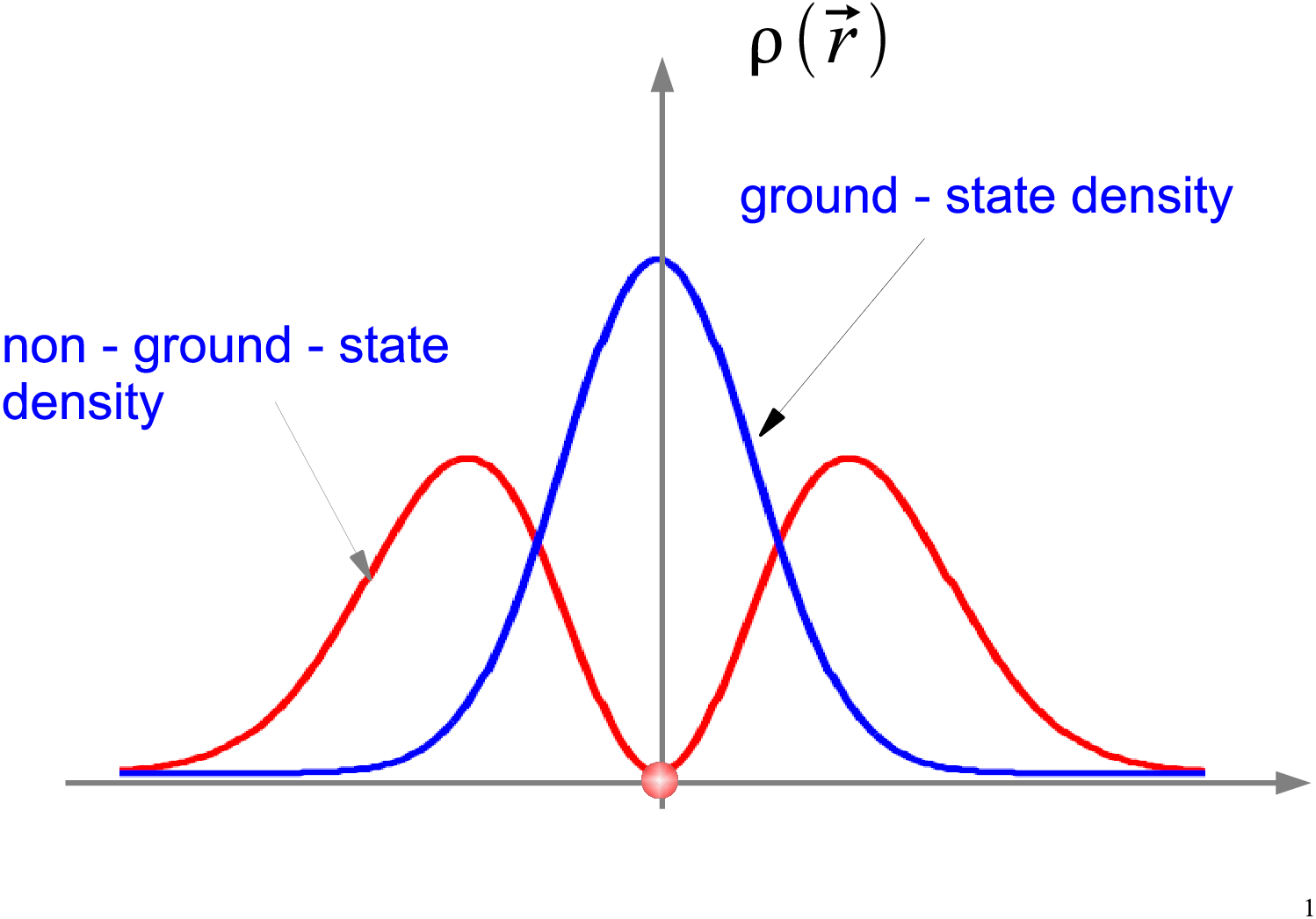}  
\end{center}
\label{Fig.3}
\end{figure}
{\bf Fig. 3}

\newpage
\begin{figure}[ht]
\begin{center}
\includegraphics[width=7cm]{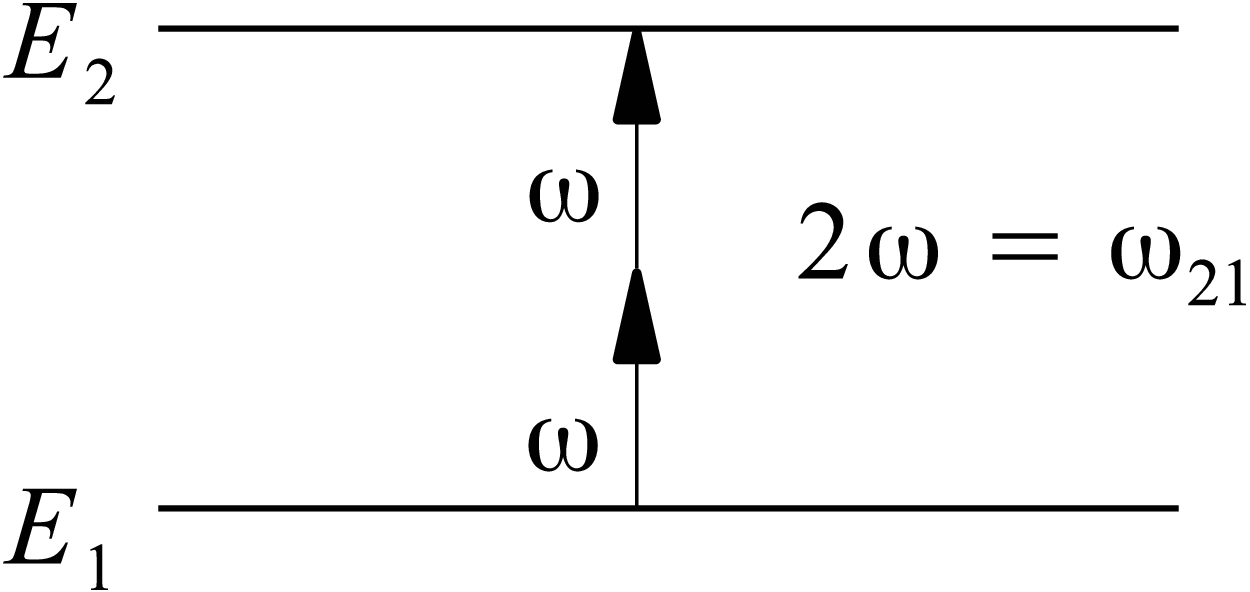}  
\end{center}
\label{Fig.4}
\end{figure}
\vskip 1.2cm
{\bf Fig. 4}

\vskip 1cm
\begin{figure}[ht]
\begin{center}
\includegraphics[width=7cm]{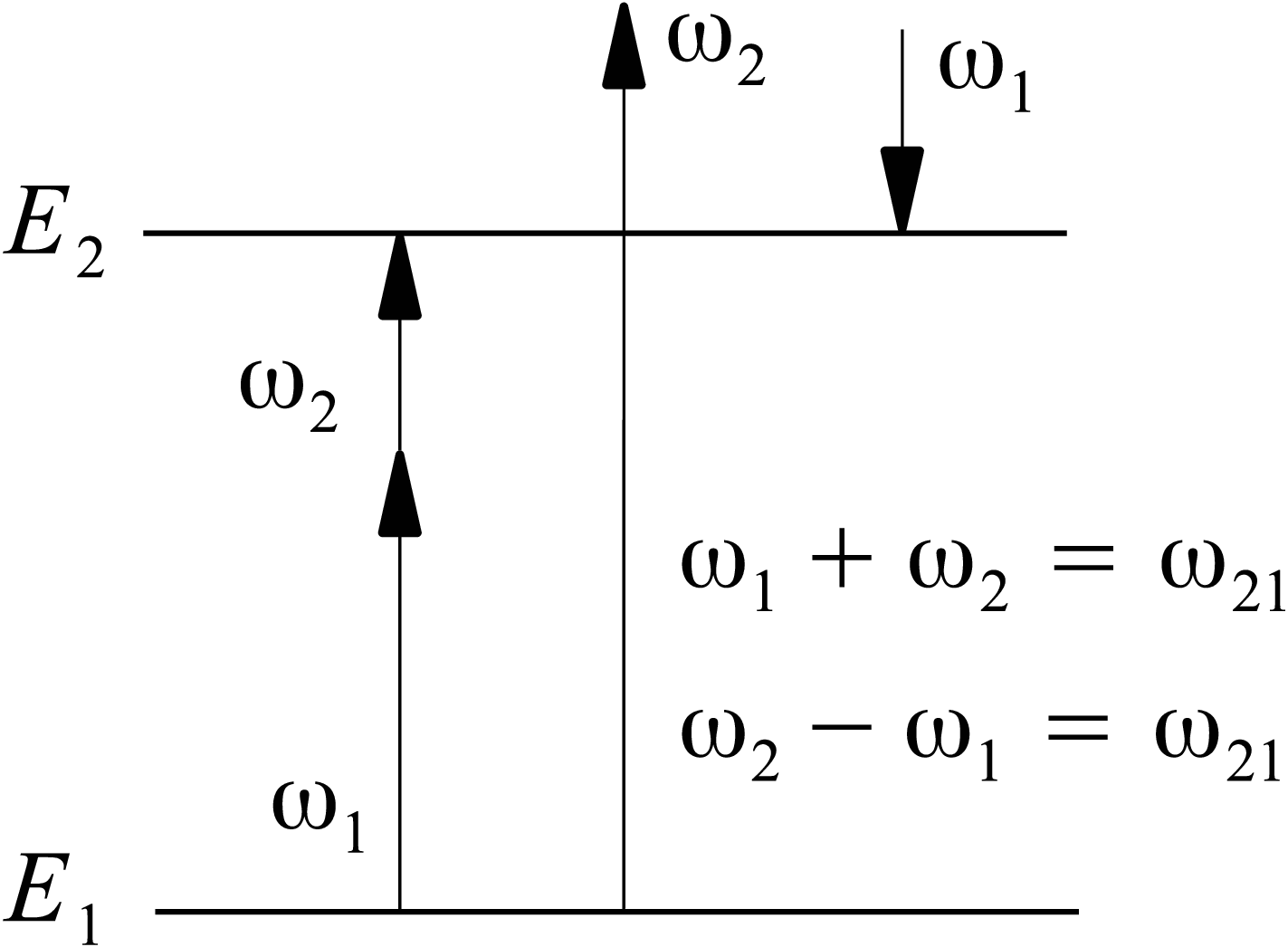}  
\end{center}
\label{Fig.5}
\end{figure}
{\bf Fig. 5}

\end{document}